\documentclass[pra,twocolumn,superscriptaddress,aps,longbibliography]{revtex4-1}
\usepackage{natbib}
\usepackage{graphicx}
\usepackage{enumitem}
\DeclareGraphicsExtensions{.pdf,.png,.jpg}
\usepackage{array} 
\usepackage{amsmath} 
\usepackage{amssymb}    
\usepackage{latexsym}
\usepackage{amsfonts}
\usepackage{mathrsfs}
\usepackage{color}
\usepackage{bbold}

\usepackage{subfigure}
\usepackage[colorlinks=true, urlcolor=blue,citecolor=blue,linkcolor=blue]{hyperref}

\begin{document}
\newcommand{\bra}[1]{\left< #1\right|}   
\newcommand{\ket}[1]{\left|#1\right>}
\newcommand{\abs}[1]{\left|#1\right|}
\newcommand{\ave}[1]{\left<#1\right>}
\newcommand{\Tr}{\mbox{Tr}}
\renewcommand{\d}[1]{\ensuremath{\operatorname{d}\!{#1}}}

\title{Operational quasiprobabilities for continuous variables}
\author{Jeongwoo Jae}
\affiliation{Department of Physics, Hanyang University, Seoul, 133-791, Republic of Korea}

\author{Junghee Ryu}
\email{rjhui82@gmail.com}
\affiliation{Centre for Quantum Technologies, National University of Singapore, 3 science Drive 2, 117543 Singapore, Singapore}

\author{Jinhyoung Lee}
\email{hyoung@hanyang.ac.kr}
\affiliation{Department of Physics, Hanyang University, Seoul, 133-791, Republic of Korea}
\begin{abstract}
 We generalize the operational quasiprobability involving sequential measurements proposed by Ryu {\em et al.} [Phys. Rev. A {\bf 88}, 052123] to a continuous-variable system. The quasiprobabilities in quantum optics are incommensurate, i.e., they represent a given physical observation in different mathematical forms from their classical counterparts, making it difficult to operationally interpret their negative values. Our operational quasiprobability is {\em commensurate}, enabling one to compare quantum and classical statistics on the same footing. We show that the operational quasiprobability can be negative against the hypothesis of macrorealism for various states of light. Quadrature variables of light are our examples of continuous variables. We also compare our approach to the Glauber-Sudarshan $\mathcal{P}$ function. In addition, we suggest an experimental scheme to sequentially measure the quadrature variables of light.
\end{abstract}

\maketitle
\section{Introduction}
Quasiprobabilities represent quantum states as phase-space distributions~\cite{Wigner32,Husimi40,Glauber63,*Sudarshan63,Cahill69,Zhang2012}. In quantum theory, the incompatible conjugate variables cannot be jointly and exactly determined due to their uncertainty relation, so the quasiprobabilities can have negative values, which is not allowed in the probability axioms~\cite{Kolmogorov33}. Thus, the negativity has been considered as a nonclassical feature of quantum systems against the classical phase-space distributions, which are always non-negative. However, this differs from other works which define nonclassicality based on the  {\em operational formalism} wherein preparation, operation, and measurement cooperate explicitly ~\cite{Spekkens05,Ferrie11}. The nonclassicality is identified by comparing the classical predictions of classical electromagnetism~\cite{Mandel65,*Mandel86} and of the realistic models~\cite{Bell64,Leggett85} assuming physical quantities are predetermined before the actual measurements. There also have been efforts to employ {\em negative probability} as criteria of describing quantum predictions~\cite{Spekkens08,Ferrie08,Ferrie12,Ryu2013,Schack00,Halliwell13,Halliwell16}.

The quasiprobabilities such as Wigner function and their classical counterparts represent a given physical observation in different mathematical forms. Furthermore, negative values in one quasiprobability can be positive in another~\cite{Ferrie11}. These may be regarded as obstacles in operationally interpreting the negative values. The former is called the ``incommensurability" of quasiprobabilities~\cite{Ryu2013}.

The commensurate approach to defining quasiprobabilities by Ryu {\em et al.}~\cite{Ryu2013} was suggested to directly compare quantum statistics to classical probability distributions in given experimental scenarios, including temporally or spatially separated observers sharing a quantum system. The quasiprobabilities are defined operationally, and called ``operational quasiprobabilities." They showed  that the negative values of the quasiprobability are incompatible with the predictions of the classical model. They considered discrete variable systems only and the generalization to continuous-variable (CV) systems has not been made yet.

In this work, we generalize the approach of operational quasiprobabilities~\cite{Ryu2013} to CV systems. A Hermite polynomial is employed to handle the unbound CV outcomes and to characterize their probability densities. We define the commensurate quasiprobabilities involving the sequential CV measurements. They consist of expectation values of unbounded observables measured at different times. Quadrature variables of light are our examples of CV systems. We prove that the existence of an underlying classical model, assuming classical realism and noninvasive measurability, called macrorealism, implies the positivity of the quasiprobabilities. The condition of no signaling in time~\cite{Kofler13,Clemente15,Clemente16} is considered a specific noninvasive measurability.

To test macrorealism, the Leggett-Garg inequality has been employed, consisting of temporal correlations between bounded variables \cite{Leggett85,Budroni14}. In this case, the observables need to be binned to dichotomic outcomes or bounded in the finite range, say, the interval~$[-1,1]$ in the macrorealism tests of CV systems~\cite{Leshem09,Asadian14,Martin16}. In Ref.~\cite{Clemente15}, unbounded observables for CV were considered to test the condition of no signaling in time; however the experimental scheme is not known yet. We propose an experimental scheme to realize the sequential CV measurements of quasiprobabilities.

We also discuss the relation of the negativity of operational quasiprobability to the nonclassicality of light, typically witnessed with the Glauber-Sudarshan $\mathcal{P}$ function~\cite{Glauber63,*Sudarshan63}. In a conventional view, coherent states and their statistical mixtures such as thermal states are understood as classical~\cite{Glauber63,*Sudarshan63,Hillery1985}, whereas states whose average photon numbers are low or superposed coherent states are nonclassical~\cite{Mandel86,Ourjoumtsev2007}. In contrast, our approach shows that the states such as vacuum, coherent, number, squeezed vacuum, Schr\"{o}dinger cat-like and a thermal state of low average photon number can have negative values in their operational quasiprobability. On the other hand, that for a bright thermal state is non-negative in the limit of an infinite number of average photons and furthermore it converges to the domain of no signaling in time.

\section{Generalization to continuous variables}
\subsection{Commensurate distribution}
A quasiprobability distribution $\mathcal{W}(x,p)$ of a quantum model is said to be {\em commensurate} with its classical counterpart of probability distribution $P(x,p)$ if both models allow the same physical interpretations for their expectations in the given functional forms. For instance, consider expectations of quantum and classical models in a given functional form of
\begin{eqnarray*}
\left \langle f(x, p) \right \rangle_Q &=& \int dx \, dp \, f(x, p) \, \mathcal{W}(x,p), \\
\left \langle f(x, p) \right \rangle_C &=& \int dx \, dp \, f(x, p) \, P(x,p).
\end{eqnarray*}
The expectation $\left \langle f(x, p) \right \rangle$ represents the statistical average of $f(x, p)$, as position $x$ and momentum $p$ are measured, if it is equated with the experimental average, i.e.,
\begin{eqnarray}
\label{eq:thwex}
\left \langle f(x, p) \right \rangle = \lim_{N \rightarrow \infty} {1 \over N} \sum_{i=1}^N f(x_i, p_i),
\end{eqnarray}
where $x_i$ and $p_i$ are the $i$th measured position and momentum.
This is exactly how the classical model interprets the functional. On the other hand, the interpretation in the quantum model depends on how one defines a quasiprobability distribution. The conventional quasiprobabilities such as Wigner, ${\cal P}$, and ${\cal Q}$ functions~\cite{Wigner32,Husimi40,Glauber63,*Sudarshan63,Cahill69} are not commensurate, as they demand different physical interpretations for a given form of functionals, not satisfying Eq.~(\ref{eq:thwex}). We need a commensurate quasiprobability to directly compare a quantum model to its classical counterpart---in other words, to keep the physical interpretation from being altered. The two types of distributions are said to be compatible when the quasiprobability distribution $\mathcal{W}(x,p)$ is non-negative everywhere in the phase space.

In quantum physics one may find it crucial to describe a quantum system in an operational way, i.e., by accounting for the physical processes in preparing, operating, and measuring a quantum state. We adopt such an operational approach in this work, contrary to the conventional approach of using mathematical transformations, e.g., Wigner-Weyl transformations, from a quantum state only. We find such a quasiprobability, in particular, satisfying the commensurability.  It was reported that a commensurate quasiprobability for a discrete variable system can be defined in an operational way~\cite{Ryu2013}.

We find such a quasiprobability distribution for a CV system, calling it an operational quasiprobability for continuous variables (OQCV). It is defined operationally with sequential and selective measurements in time. However, before doing so, it is mathematically convenient to expand arbitrary normalized distributions in terms of Hermite polynomials. For instance, consider and expand a probability distribution of two arguments,
\begin{equation}
\label{eq:C}
P(x_1,x_2) = e^{-\frac{1}{2}x_1^2}e^{-\frac{1}{2}x_2^2}\sum_{m,n=0}^{\infty}\frac{H_m(x_1)H_n(x_2)}{2\pi m!n!} C_{mn},
\end{equation}
where $H_n(x)$ is a Hermite polynomial of the $n$th degree. Reciprocally,
\begin{eqnarray}
\label{eq:charac}
C_{mn}&=&\int\int dx_1 dx_2 \, H_m(x_1)H_n(x_2) \, P(x_1,x_2) \\
&=& \left \langle H_m(x_1)H_n(x_2) \right \rangle, \nonumber
\end{eqnarray}
for $m, n = 0, 1, 2, ...$. It is seen that $C_{mn}$ contains the complete information on the distribution $P(x_1,x_2)$, called a characteristic tensor. Here we used the fact that Hermite polynomials form a complete set of orthogonal bases in the real space with respect to the weight function $\text{exp}(-x^2/2)$~\cite{Szego39}:
\begin{equation}\label{eq:ortho}
\int dx~ e^{-\frac{1}{2}x^2} H_k(x)H_l(x) = \sqrt{2\pi}k!\delta_{kl}.
\end{equation}
Note that the integral is taken over $(-\infty,+\infty)$; this convention is used below in all formulas. For more detailed calculation, see Appendix \ref{eq:character}.
\begin{figure}[t]
	\centering
	\includegraphics[width=0.5\textwidth]{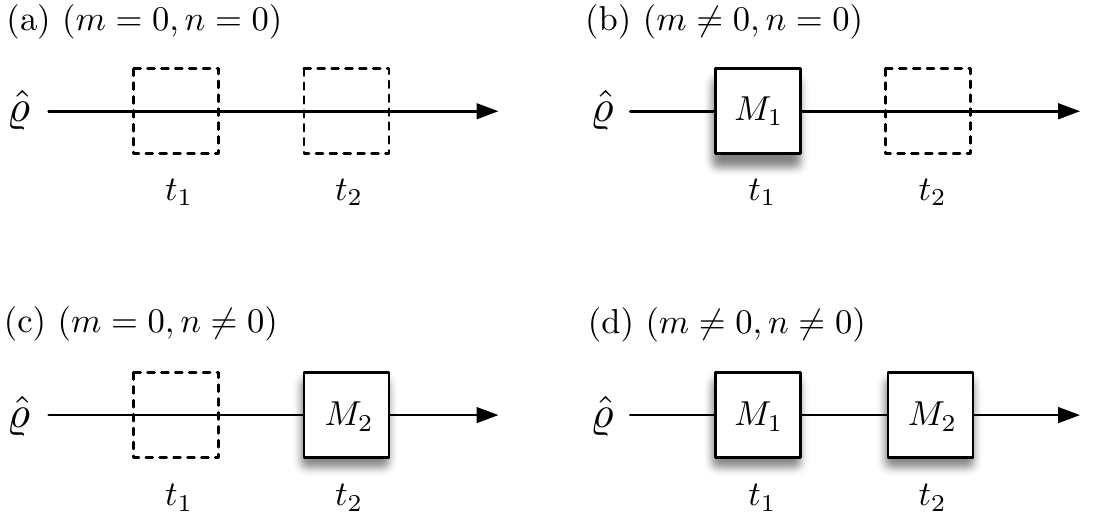}
	\caption{Measurement setups when two observables are considered. On implementing the measurements, four cases are possible: (a) the reference of no measurement, (b), (c) the single measurements of the two alternative observables, and (d) the sequential measurement of them. Each setup is denoted by the tuple $(m,n)$, which is associated with the degrees of Hermite polynomials $H_{m} (x)$ and $H_{n} (x)$.}
	\label{fig:setting}
\end{figure}

\subsection{Selective sequential measurement}
We shall define OQCV with two sequential and selective measurements in time. Suppose that two measurements $M_i$ for $i \in \{1,2\}$ are selected to be performed at time $t_i$ (here $t_1<t_2$), and their outcomes $x_i$ are real numbers, i.e., $x_i\in(-\infty,+\infty)$. There are four cases possible (see Fig. \ref{fig:setting}): [Fig.~\ref{fig:setting}(a)] the reference of no measurement, [Figs.~\ref{fig:setting}(b) and (c)] the single alternative measurements, and [Fig.~\ref{fig:setting}(d)] both measurements performed sequentially. Here we denote the four setups by the tuple $(m, n)$, where $m\neq0$ (or $n\neq 0$) implies that the respective measurement $M_1$ (or $M_2$) is to be performed. For each measurement setup, we obtain the expectations of
\begin{eqnarray}
\Gamma_{00} &=& 1, \nonumber \\
\Gamma_{m0} &=&  \int dx_1 H_m(x_1)P(x_1|M_1)~~ \text{for}~ m\neq 0,  \nonumber \\
\Gamma_{0n} &=&  \int dx_2 H_n(x_2)P(x_2|M_2)~~ \text{for}~ n\neq 0,  \nonumber \\
\Gamma_{mn} &=&  \int\int dx_1dx_2 H_m(x_1)H_n(x_2)P(x_1,x_2|M_1,M_2)\nonumber \\
&&\text{for}~ m,n\neq 0,
\label{eq:gamma}
\end{eqnarray}
where $P(x_i|M_i)$ are experimental probability distributions in measurements $M_i$ and $P(x_1,x_2|M_1,M_2)$ is in the joint measurement. Here, $\Gamma$s are the moments of Hermite polynomials in the measurement setups, respectively.

We now propose the OQCV defined by
\begin{equation}
\mathcal{W}(x_1,x_2) \equiv e^{\text{-}\frac{1}{2}x_1^2}e^{\text{-}\frac{1}{2}x_2^2}\sum_{m,n=0}^{\infty}\frac{\Gamma_{mn}}{{2\pi} m!n!} H_m(x_1)H_n(x_2).
\label{eq:quasi}
\end{equation}
It can be represented in terms of the (experimental) probabilities (see Appendix \ref{eq:deriving}):
\begin{eqnarray}
\begin{aligned}
\mathcal{W}&(x_1,x_2) = P(x_1,x_2|M_1,M_2)  \\
&+ \frac{1}{\sqrt{2\pi}}e^{-\frac{1}{2}{x^2_2}}\left[ P(x_1|M_1)-P(x_1|M_1,M_2) \right] \\
&+ \frac{1}{\sqrt{2\pi}}e^{-\frac{1}{2}{x^2_1}}\left[ P(x_2|M_2)-P(x_2|M_1,M_2) \right].
\label{eq:w}
\end{aligned}
\end{eqnarray}
It is worth noting that $\mathcal{W}(x_1, x_2)$ has the following properties:

\subsubsection{{\em Commensurability}} 
The $\mathcal{W}(x_1, x_2)$ can represent the characteristic tensor
\begin{equation}\label{eq:characterisitc}
\Gamma_{mn}=\int\int dx_1 dx_2H_m(x_1)H_n(x_2)\mathcal{W}(x_1,x_2),
\end{equation}
for all non-negative integers of $(m,n)$. It implies that the distribution $\mathcal{W}(x_1, x_2)$ governs the statistics of the four measurement setups in Fig.~\ref{fig:setting}.

\subsubsection{{\em Marginality}} The marginal of $\mathcal{W}(x_1, x_2)$ for a variable $x_1$ becomes the probability distribution of measuring $x_2$, i.e., $\int dx_1 ~\mathcal{W}(x_1,x_2) = P(x_2|M_2)$, the same as for $x_2$. Accordingly, the $\mathcal{W}(x_1, x_2)$ is normalized as $\int \int dx_1dx_2 ~\mathcal{W}(x_1,x_2) =1$.

\subsection{Two theoretical models of OQCV}
A classical prediction of the OQCV is determined by the hypotheses according to a physical circumstance. We consider a classical model assuming realism and noninvasive measurability. Classical physics has been considered as the realistic theory which assumes predetermined physical quantities before the actual measurements. This implies the existence of an underlying joint probability distribution for the outcomes of all possible measurements.

In a temporal scenario, Leggett and Garg examined noninvasive measurability at the macroscopic level. One can measure a physical quantity of a macroscopic object without disturbing it. This hypothesis together with realism, called macrorealism (MR), leads the Leggett-Garg inequality involving temporal correlations~\cite{Leggett85}. It shows that quantum prediction is incompatible with the classical one. More precisely, MR is defined by the following three hypotheses~\cite{Leggett02,Kofler08}: ``{\em Macrorealism per se}. A macroscopic object which has available to it two or more macroscopically distinct states is at any given time in a definite one of those states. {\em Non-invasive measurability} (NIM). It is possible in principle to determine which of these states the system is in without any effect on the state itself or on the subsequent system dynamics. {\em Induction} [also called arrow of time (AoT)]. The properties of ensembles are determined exclusively by initial conditions".

Recently, the no-signaling-in-time (NSIT) condition was suggested as a statistical version of the NIM model. It states that: ``a measurement does not change the outcome statistics of later measurement"~\cite{Kofler13}. The conjunction of NSIT and AoT is necessary and sufficient for MR~\cite{Clemente15}. We take a classical model with the hypotheses of the NIST and AoT conditions, each of which reads
\begin{eqnarray*}
\text{NSIT}&:&~\int dx_1~P(x_1,x_2|M_1,M_2)=P(x_2|M_2), \\
\text{AoT}&:&~\int dx_2~P(x_1,x_2|M_1,M_2)=P(x_1|M_1).
\end{eqnarray*}
The AoT condition is satisfied by both quantum and classical theories~\cite{Ryu2013,Halliwell16}.  

\begin{figure}[t]
	\centering
	\includegraphics[width=0.36\textwidth]{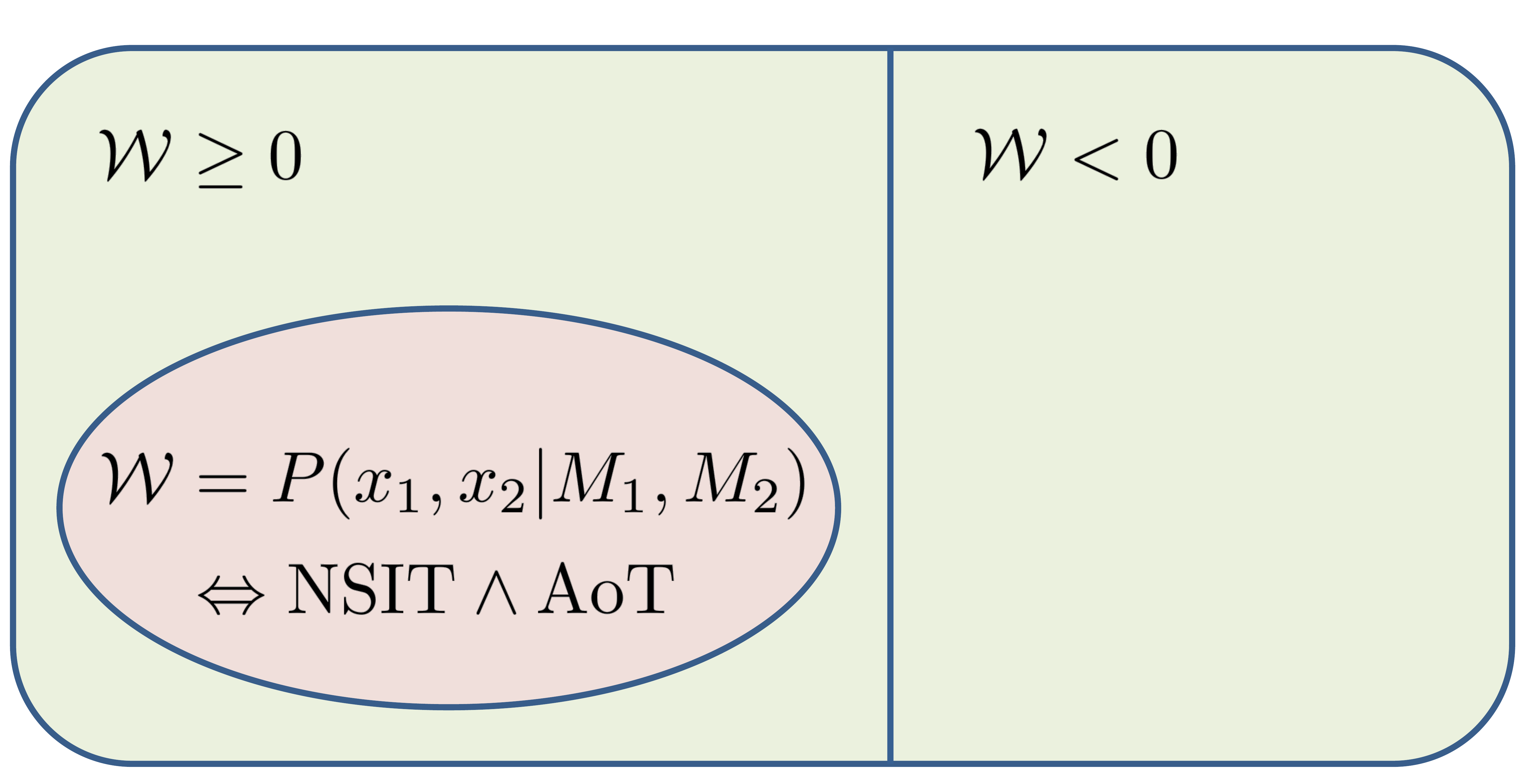}
	\caption{Venn diagram for theoretical predictions of operational quasiprobability distributions $\mathcal{W}(x_1,x_2)$ for measurements $M_1$ and $M_2$. In a classical model assuming the NSIT and AoT conditions, $\mathcal{W}$ becomes a joint probability $P(x_1,x_2|M_1,M_2)$. However, the $\mathcal{W}$ can still be non-negative even though the condition of no signaling in time is violated.}
	\label{fig:Venn}
\end{figure}

In such a classical model, the OQCV [Eq.~(\ref{eq:w})] becomes a joint probability distribution, i.e., $\text{NSIT}\wedge\text{AoT}\Leftrightarrow\mathcal{W}=P(x_1,x_2|M_1,M_2)\geq0$. We assume that the AoT condition holds. $\mathcal{W}$ can be negative depending on the degree of violation of the NSIT condition: $P(x_2|M_2)-P(x_2|M_1,M_2)$. The negativity of $\mathcal{W}$ is a sufficient condition for violating NSIT, i.e., $\mathcal{W}<0 \Rightarrow \neg \text{NSIT}$. However, the inverse does not hold. The $\mathcal{W}$ can be non-negative even though the NSIT condition is violated. The relation between $\mathcal{W}$ and the NSIT condition is depicted in Fig.~\ref{fig:Venn}.

For quantum theory, we consider a quantum state $\hat{\varrho}$ and positive operator valued measure (POVM) $\{\hat{A}^{\dagger}(x)\hat{A}(x)\}$ of outcome $x$ (they do not commute each other in general). Each probability distribution reads as follows: For a single measurement $P(x_i|M_i)=\Tr[\hat{A}(x)\hat{\varrho}\hat{A}^{\dagger}(x)]$ and for a sequence one $P(x_1,x_2|M_1,M_2)=\Tr[\hat{A}(x_2)\hat{A}(x_1)\hat{\varrho}\hat{A}^{\dagger}(x_1)\hat{A}^{\dagger}(x_2)]$. The probability $P(x_i | M_1,M_2)$ for $i \in \{1,2\}$ can be obtained marginally from $P(x_1,x_2 | M_1,M_2)$, respectively.

As a nonclassicality measure, we employ a total volume of the negative probability of the OQCV, negativity $\mathcal{N}$~\cite{Kenfack04,Ryu2013}:
\begin{equation}
\mathcal{N} \equiv \frac{1}{2}\int\int dx_1dx_2 \big[\abs{\mathcal{W}(x_1,x_2)} - \mathcal{W}(x_1,x_2) \big].
\label{eq:N}
\end{equation}
To collect full contributions of the negative volume, one examines all possible measurement bases.

\section{Negativity of Quadrature variables of light}

\begin{figure}[t]
	\centering
	\includegraphics[width=0.48\textwidth]{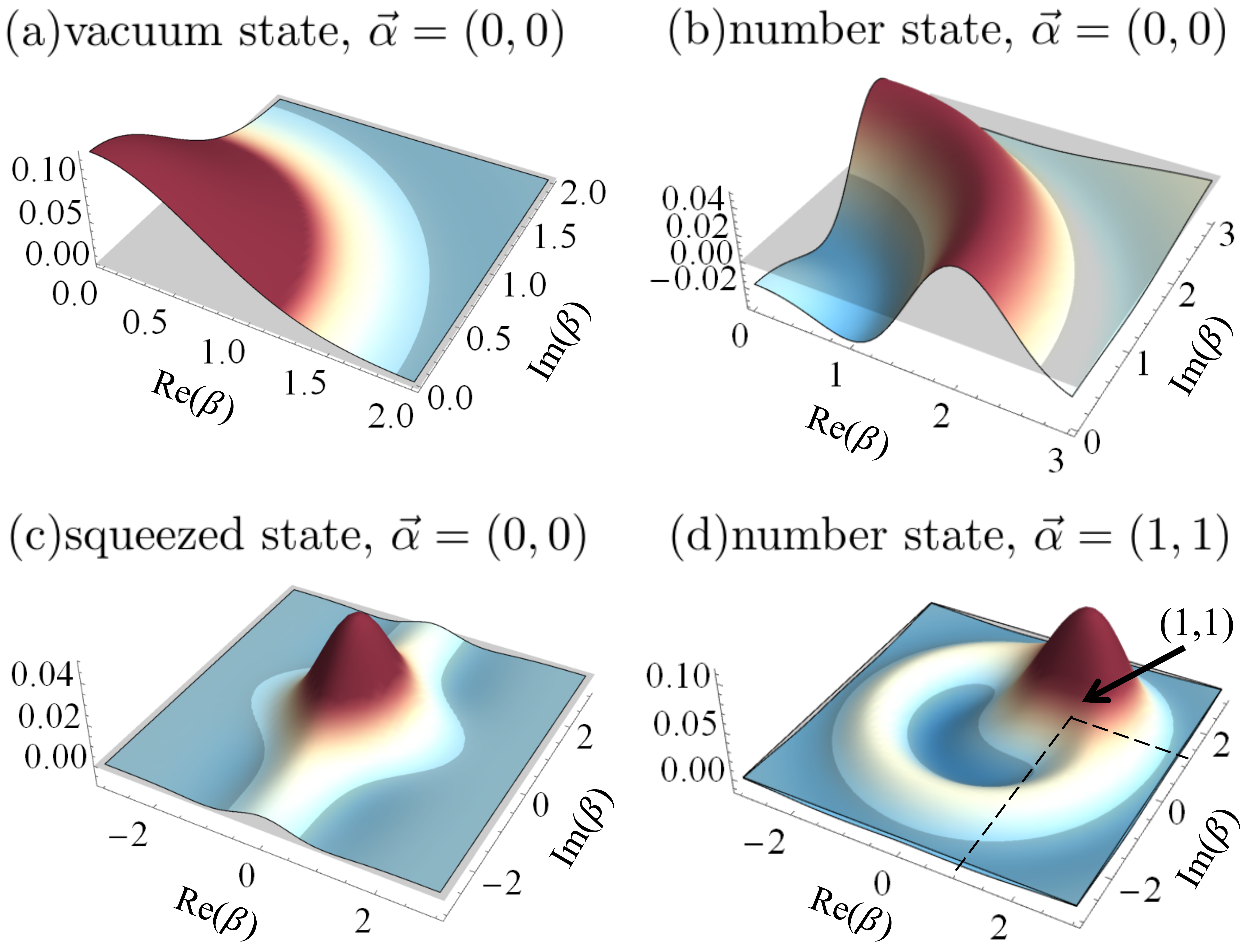}
	\caption{Operational quasiprobabilities for some states, $\mathcal{W}(\vec{\alpha},\vec{\beta})=\mathcal{W}(\alpha_r, \alpha_i, \beta_r, \beta_i)$, for fixed $\vec{\alpha}$. The gray plane implies $\mathcal{W}=0$, and the blue regions (below the plane) are negative values of the $\mathcal{W}$. We plot $\mathcal{W}(\vec{0},\vec{\beta})$ for (a) the vacuum, (b) the number state $\ket{2}$, and (c) the squeezed state of average photon number $\bar{n}_{\text{sq}}=5$. The plots of (a) and (b) are symmetric in the azimuthal direction. In (d), we plot $\mathcal{W}$ of number state $\ket{2}$ is plotted by changing the $M_1$ basis from $\vec{\alpha}=(0,0)$ to $(1,1)$. The positive (red) regions in turn appear around $\vec{\beta}=(1,1)$.}
	\label{fig:Wcoherent}
\end{figure}

We examine OQCV for the quadrature variables of light. As our examples of light, we consider vacuum, coherent, number, squeezed vacuum, cat, and (bright) thermal states. It turns out that the negativity depends on the overlap between the given state and measurement bases on the phase-space. To see this more clearly, we plot OQCV for some states with a fixed measurement basis, which is presented in Fig.~\ref{fig:Wcoherent}. We additionally analyze how the overlap contributes to the negativity for the example of a coherent state. For each considered state, we numerically evaluate the negativity as a function of average photon number, which is presented in Fig.~\ref{fig:negativity}. Furthermore, an operational meaning of the OQCV's negativity is discussed with respect to the Glauber-Sudarshan $\mathcal{P}$ function.

\subsection{OQCV for quadrature variables}

The distribution of quadrature variables, called the Husimi $\mathcal{Q}$ function~\cite{Husimi40}, is obtained by the coherent state basis measurement $\pi^{\text{-}1}|\alpha\rangle\langle\alpha|$. It is a POVM satisfying overcompleteness $\pi^{\text{-}1}\int d^2\alpha|\alpha\rangle\langle\alpha|=\mathbb{1}$, where $d^2\alpha$ stands for $d\alpha_r d\alpha_i$. Thus, the measurement outcome is obtained in the form of two real numbers in the pair $\alpha=\alpha_r+i\alpha_i$, and thus we use vector $\vec{\alpha} = (\alpha_r, \alpha_i)$ to represent the each measurement outcome.

The OQCV for a light state $\hat{\varrho}$, with the measurement bases $\pi^{\text{-}1}|\alpha\rangle\langle\alpha|$ of $M_1$ and $\pi^{\text{-}1}|\beta\rangle\langle\beta|$ of $M_2$, is given by
\begin{equation}
\begin{aligned}
\mathcal{W}&(\vec{\alpha},\vec{\beta}) = P(\vec{\alpha},\vec{\beta}|M_1,M_2) \\
&+ \frac{1}{2\pi}e^{-\frac{1}{2}\abs{\alpha}^2}\left[ P(\vec{\beta}|M_2)-P(\vec{\beta}|M_1,M_2) \right],
\label{eq:quadrature}
\end{aligned}
\end{equation}
where $\mathcal{W}(\vec{\alpha},\vec{\beta})=\mathcal{W}(\alpha_r, \alpha_i, \beta_r, \beta_i)$. Here, $P(\vec{\alpha}|M_1)=\pi^{\text{-}1}\langle\alpha |\hat{\varrho}|\alpha\rangle$, $P(\vec{\beta}|M_1)=\pi^{\text{-}1}\langle\beta |\hat{\varrho}|\beta\rangle$ and $P(\vec{\alpha}, \vec{\beta}|M_1,M_2)=\pi^{\text{-}2}\abs{\langle\beta |\alpha\rangle}^2\langle\alpha |\hat{\varrho}|\alpha\rangle$. We note that the term $P(\vec{\alpha}|M_1)-P(\vec{\alpha}|M_1,M_2)$ disappears as the AoT condition holds in the quantum model. The four Hermite polynomials are employed to transform the two-pair continuous variables $(\alpha_r, \alpha_i)$ and $(\beta_r, \beta_i)$ (refer to the paragraph below the equation in Appendix~\ref{eq:deriving}). We consider vacuum $|0\rangle$, coherent state $|w\rangle$, number state $|n\rangle$, squeezed vacuum state $|s_r\rangle$ with the squeezing $\abs{r}$, cat state ${A}_{\pm}^{\text{-}1}(\ket{w}\pm\ket{-w})$ with normalization factor ${A}_{\pm}={[2\pm 2\exp({-2\abs{w}^2})]}^{1/2}$, and thermal state $\sum_{n=0}^{\infty}{{(\bar{n}_{\text{th}})}^{n}}/{(\bar{n}_{\text{th}}+1)^{n+1}}|n\rangle\langle{n}|$ with average photon number $\bar{n}_{\text{th}}$. For each considered state, the average photon numbers are $\bar{n}_{\text{vac}}=0$ for vacuum, $\bar{n}_{\text{co}}=\abs{w}^2$ for the coherent state, $\bar{n}_{\text{num}}=n$ for the number state, $\bar{n}_{\text{sq}}=\sinh^2 \abs{r}$ for the squeezed vacuum state, $\bar{n}_{+}=\abs{w}^2\tanh\abs{w}^2$ for the plus cat state, and $\bar{n}_{-}=\abs{w}^2\coth\abs{w}^2$ for the minus cat state.

\subsection{Results}
As we pointed out before, the negativity of OQCV is determined by the overlap between the given state and the measurement bases. In Fig.~\ref{fig:Wcoherent}, we plot the OQCV as a function of $\vec{\beta}$ by fixing $\vec{\alpha}$ (in general, the OQCV lives in four-dimensional space; thus we fix one measurement basis for plots). We consider the vacuum, number state $\ket{2}$, and squeezed vacuum states of $\bar{n}_{\text{sq}}=5$. The blue regions of each plot denote the negative values of the OQCV. We also observe the behavior of moving the positive (red) regions of the OQCV for number state $\ket{2}$ as changing the $M_1$ basis from $\vec{\alpha}=(0,0)$ to $(1,1)$, [see Fig. \ref{fig:Wcoherent}(d)].

Let us now examine how the overlap contributes to the negativity of the OQCV for a coherent state $|w\rangle$. The OQCV reads
\begin{equation}
\begin{aligned}
\label{eq:overlap}
\mathcal{W}&_{\ket{w}}(\vec{\alpha},\vec{\beta})=\frac{1}{\pi^2}\abs{\langle w|\alpha\rangle}^2\abs{\langle\alpha|\beta\rangle}^2 \\
&+\frac{1}{2\pi}e^{-\frac{1}{2}\abs{\alpha}^2}\left( \frac{1}{\pi}\abs{\langle w|\beta\rangle}^2-\frac{1}{2\pi}\abs{\langle w|\beta\rangle}\right),
\end{aligned}
\end{equation}
where the second term in the bracket, the marginal probability $P(\vec{\beta}|M_1,M_2)$ in Eq.~(\ref{eq:quadrature}), was obtained marginally from the joint probability distribution $P(\vec{\alpha},\vec{\beta}|M_1,M_2)$. That is, $(2\pi)^{-1}e^{-\frac{1}{2}\abs{w-\beta}^2}=(2\pi)^{-1}\abs{\langle w|\beta\rangle}$.

The negativity of $\mathcal{W}_{\ket{w}}$ is determined by the difference in the bracket in Eq.~(\ref{eq:overlap}) as the first term represents the joint probability distribution; i.e., it is always positive semidefinite. The difference in the bracket $\pi^{\text{-}1}\abs{\langle\beta|w\rangle}(\abs{\langle\beta |w\rangle}-1/2 )$ is non-negative if $\abs{\langle\beta|w\rangle}\geq1/2$. In the case of $\abs{\langle w|\beta\rangle}< 1/2$, $\mathcal{W}_{\ket{w}}$ is negative in the competition between the negative difference and the positive joint probability. In particular, the value of the bracket in Eq.~(\ref{eq:overlap}) is minimized when $\abs{\langle w|\beta\rangle}=1/4$, so that the OQCV can be negative when the first measurement basis $\vec{\alpha}$ satisfies $\abs{\langle w|\alpha\rangle}^2\abs{\langle \alpha |\beta\rangle}^2 < 2^{-5}\abs{\langle0 |\alpha\rangle}$.

\begin{figure}[t]
	\centering
	\includegraphics[width=0.48\textwidth]{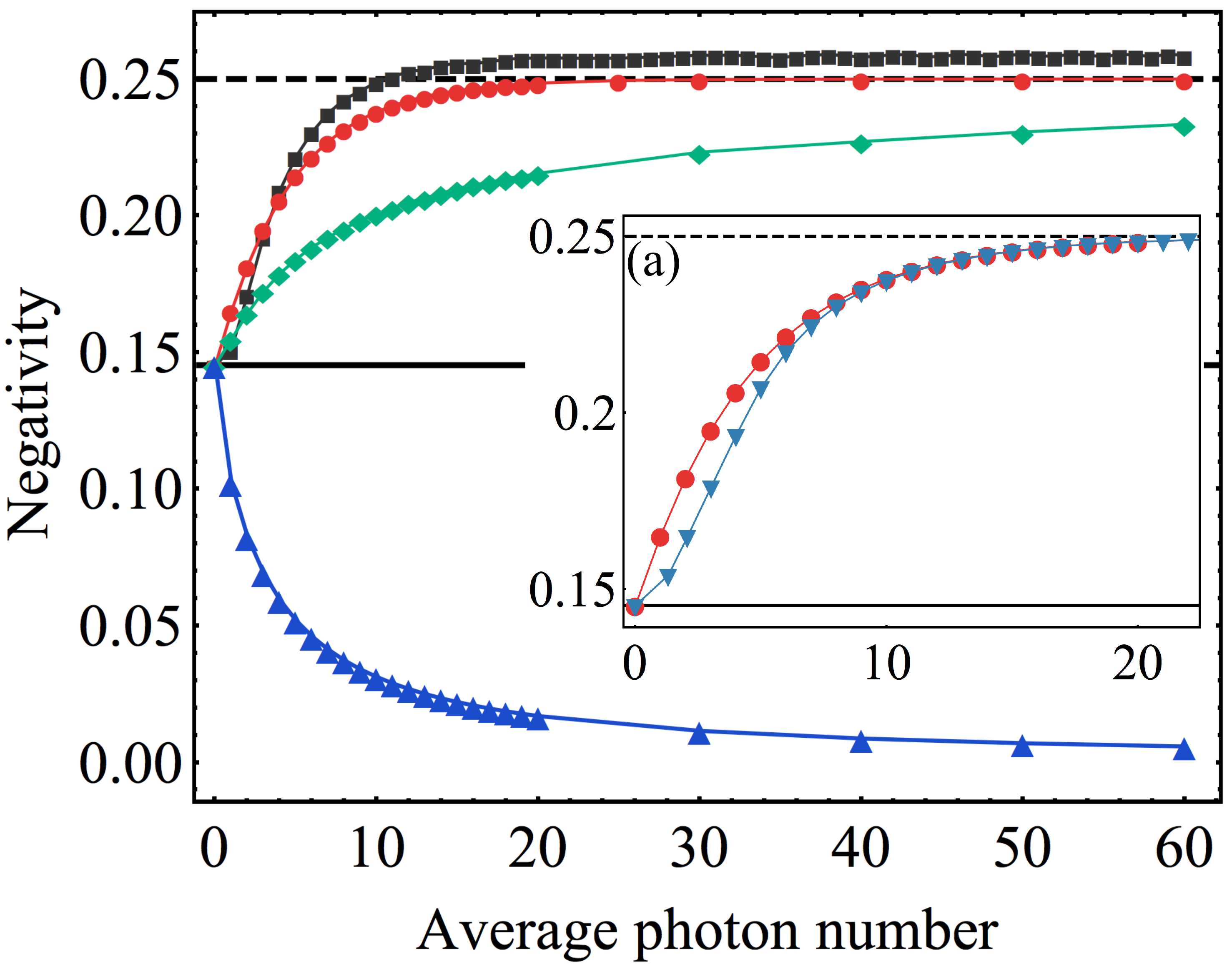}
	\caption{Negativity $\mathcal{N}$ for the vacuum, coherent ($\bullet$), number (${\blacksquare}$), squeezed vacuum ($\blacklozenge$), plus and minus cat states ($\blacktriangledown$), and thermal state ($\blacktriangle$) as a function of average photon number $\bar{n}$. The $\mathcal{N}$ of coherent, squeezed vacuum, and cat states is saturated around $\bar{n}= 20$, and in particular the coherent state is saturated to $\mathcal{N}=0.250045$, denoted by dashed line. For the vacuum state $\bar{n}=0$, $\mathcal{N}=0.145420$ (solid line). In contrast, for the thermal state, $\mathcal{N}$ is decreasing as $\bar{n}$ is increasing. (a) The negativitities of plus and minus cat states are the same, and they are close to that of the coherent state around $\bar{n}=10$.}
	\label{fig:negativity}
\end{figure}

Figure~\ref{fig:negativity} shows numerical results of the negativity of various quantum states. We plot the negativity $\mathcal{N}$ as a function of average photon number. It turns out that all considered states (except the bright thermal state) are nonclassical, i.e., they all have negative values in their OQCV. We observe that of the coherent, squeezed vacuum, and cat states have the negativity that saturates to $\bar{n}\approx 20$; in particular the coherent state saturates to $\mathcal{N_{\text{co}}} = 0.250045$. For the squeezed vacuum state, its negativity still increases persistently. The vacuum state shows $\mathcal{N}_{\text{vac}}=0.145420$. In contrast, the thermal state shows the opposite behavior from the others. The negativity of the thermal state, $\mathcal{N}_{\text{th}}$, decreases as the average photon number $\bar{n}_{\text{th}}$ increases, and finally $\mathcal{N}_{\text{th}} \rightarrow 0$ as $\bar{n}_{\text{th}} \rightarrow \infty$. Furthermore, it is remarkable that the bright thermal state converges to the domain of no signaling in time and thus its limit is classical (see Appendix \ref{SEC:APP_C} for more details).

\subsection{Arbitrary state in Glauber-Sudarshan $\mathcal{P}$ representation}\label{section:op}
An arbitrary light state $\hat{\varrho}$ can be represented in terms of the coherent state basis $|w\rangle\langle w|$ with the Glauber-Sudarshan $\mathcal{P}$ function~\cite{Glauber63,*Sudarshan63}:
\begin{equation}
\hat{\varrho}=\int d^2 w \mathcal{P}_{\varrho}(w)|w\rangle\langle w|.\nonumber
\end{equation}
The state $\hat{\varrho}$ is said to be nonclassical if the $\mathcal{P}$ function is negative or highly singular~\cite{Mandel86,Kiesel10,Note1}. The OQCV of arbitrary $\hat{\varrho}$ is then given by the expectation of the $\mathcal{W}_{\ket{w}}$ over the $\mathcal{P}$ function:
\begin{equation}
\mathcal{W}_{\hat{\varrho}}(\vec{\alpha},\vec{\beta}) =\int d^2 w \mathcal{P}_{\varrho}(w) ~\mathcal{W}_{\ket{w}}(\vec{\alpha},\vec{\beta}).\nonumber
\end{equation}
It is worth noting that operationally defined OQCV reveals the negativity by an interplay of a given state and measurements. The state $\hat{\varrho}$ is considered by OQCV to be nonclassical if, for some $\vec{\alpha}$, and $\vec{\beta}$,
\begin{equation}
\int d^2 w\mathcal{P}_{\varrho}(w)\mathcal{W}_{\ket{w}}(\vec{\alpha},\vec{\beta})<0.
\end{equation}
While vacuum, coherent, and thermal states have positive $\mathcal{P}$ functions, their OQCVs can be negative.

\section{Experimental scheme}
\begin{figure}[t]
	\centering
	\includegraphics[width=0.48\textwidth]{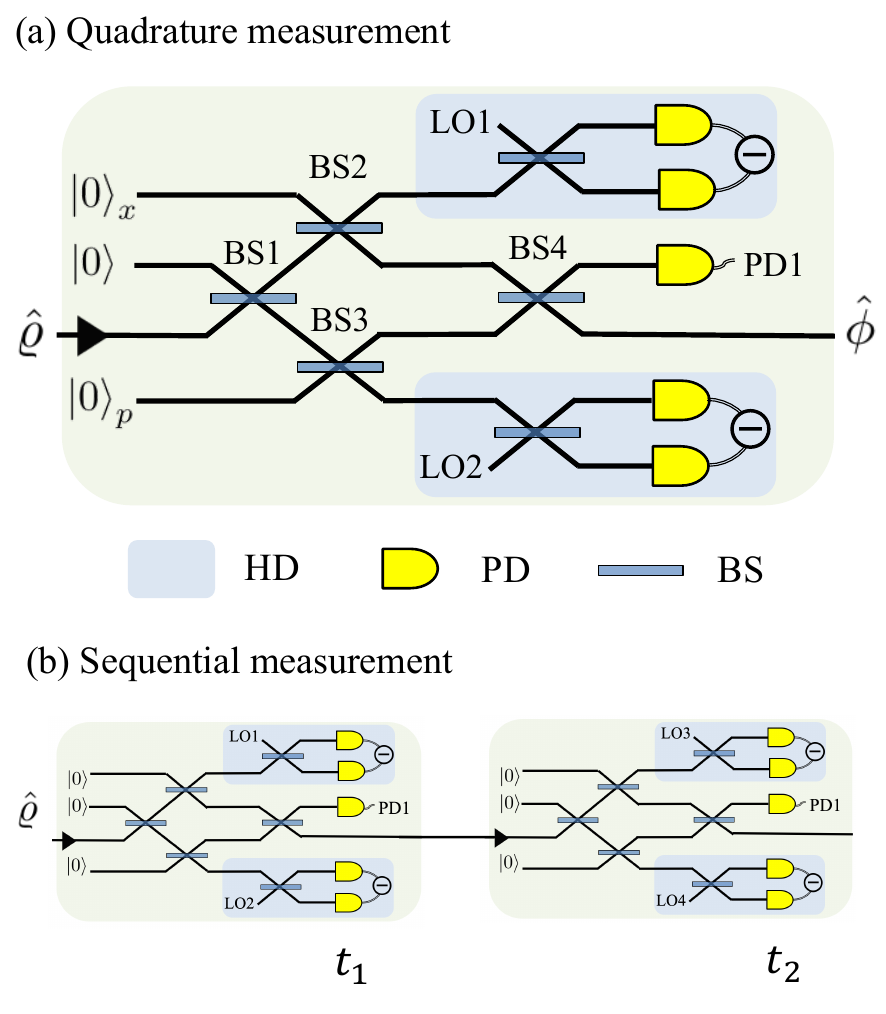}
	\caption{Optical scheme to sequentially measure quadrature variables. (a) Quadrature variables of input state $\hat{\varrho}$ are obtained by jointly measuring homodyne detection (HD). Their local oscillators (LOs) 1 and 2 are locked to measure $x$ and $p$ quadratures. We employ a beam splitter to entangle the input state with ancillary zero states $\ket{0}_{x,p}$. We consider the ancillary states to be eigenstates of zero quadrature, i.e., $\hat{x}\ket{0}_x=0$ and $\hat{p}\ket{0}_p=0$. The obtained statistics is a scaled Husimi $\cal{Q}$ function ${\cal{Q}} (\sqrt{2} \mu)$ of the measurement basis $\mu=x+ip$. The postmeasurement state $\hat{\phi}$ is obtained as the coherent state $|\mu\rangle\langle\mu |$ when photodetector (PD) 1 is not clicked. (b) Sequential measurement of the quadrature variables can be realized by performing the $\cal{Q}$-function measurement consecutively at two different times $t_1,~t_2$.}
	\label{fig:tele}
\end{figure}

We propose an experimental scheme for the sequential measurement of quadrature variables. The coherent state basis measurement is implemented by using homodyne detection~\cite{Noh92,Vogel93} and a joint scheme~\cite{Stenholm92,Leonhardt93,Welsch99} called heterodyne detection. The typical homodyne detection scheme is employed for the state tomography such that the input light totally vanishes by photocounter. To sequentially measure quadrature variables in OQCV, we demand a nonvanishing optical measurement scheme. Our experimental proposal can be applied to arbitrary states.

Our optical scheme consists of beam splitters and two homodyne detections whose local oscillators are locked to measure $(x,p)$-quadrature variables. The beam splitters (BS) transform input field $\hat{a}_1,\hat{a}_2$ to output field $\hat{a}_3,\hat{a}_4$ with the relations $\hat{B}\hat{a}_1\hat{B}^{\dagger} = (\hat{a}_3+\hat{a}_4)/\sqrt{2}$ and $\hat{B}\hat{a}_2\hat{B}^{\dagger}=(-\hat{a}_3+\hat{a}_4)/\sqrt{2}$. Two ancillary states $\ket{0}_{x,p}$ are also used to creat entanglement with the input state. The ancillary states are prepared as the zero eigenstates of recpective quadrature variables, i.e., $\hat{x}\ket{0}_x=0$, $\hat{p}\ket{0}_p=0$. The scheme is illustrated in Fig.~\ref{fig:tele}(a).

Consider a light state $\hat{\varrho}=\int d^2\alpha \mathcal{P}(\alpha)|\alpha\rangle\langle\alpha|$ as an input state. The state first passes BS1, and the output state $\hat{\varrho}'$ is $\hat{B}_1~\hat{\varrho}\otimes|0\rangle\langle0|~\hat{B}^{\dagger}_1=\int~d^2\alpha \mathcal{P}(\alpha)|\alpha/\sqrt{2}\rangle\langle\alpha/\sqrt{2}|\otimes|{\alpha}/{\sqrt{2}}\rangle\langle\alpha/\sqrt{2}|$. BS2 and BS3 are also applied and mix $\hat{\varrho}'$ with the zero eigenstates $\ket{0}_x,\ket{0}_p$. This results in an entangled state in the quadrature basis. The entangled state is given by $\hat{B}_3\hat{B}_2\big(\hat{\varrho}'\otimes |0\rangle_x\langle 0|\otimes |0\rangle_p\langle 0|)\hat{B}^{\dagger}_2\hat{B}^{\dagger}_3$, and its diagonal elements read
\begin{eqnarray}\label{eq:entangle}
&&~~~~\int d^2\alpha dxdp~\mathcal{P}(\alpha) \abs{\Psi(x)}^2 \abs{\Psi(p)}^2 \nonumber\\
&&\times \left|\frac{x}{\sqrt{2}} ~\frac{x}{\sqrt{2}} \right\rangle \left\langle \frac{x}{\sqrt{2}} ~\frac{x}{\sqrt{2}} \right|\otimes \left|\frac{p}{\sqrt{2}} ~\frac{p}{\sqrt{2}} \right \rangle \left\langle \frac{p}{\sqrt{2}}~\frac{p}{\sqrt{2}} \right|,\nonumber
\end{eqnarray}
where the $\Psi(x), \Psi(p)$ are the wave functions for the coherent state represented in $x,p$ space. In this case, $\Psi(x)=\pi^{-1/4}e^{{-\frac{1}{2}(x-\alpha_r)^2+i\alpha_ix-i\alpha_r\alpha_i/2}}$ and $\Psi(p)=\pi^{-1/4}e^{{-\frac{1}{2}(p-\alpha_i)^2-i\alpha_rp+i\alpha_r\alpha_i/2}}$ for the coherent state $\ket{\alpha/\sqrt{2}}=\ket{(\alpha_r+i\alpha_i)/\sqrt{2}}$. We only consider the diagonal elements of the state as the quadrature measurements $|x \rangle\langle x |$, $|p \rangle\langle p |$ will be performed at the ancillary modes. 

To obtain the entangled state, we use the fact that coherent states and zero eigenstates can be entangled when they are mixed via BS~\cite{Samuel98,Parker2000}:
\begin{eqnarray}
\hat{B}\ket{\frac{\alpha}{\sqrt{2}}}\ket{0}_x &=& \hat{B}\int dxdx' \Psi(x)\delta(x')\ket{x}\ket{x'} \nonumber\\
&=&\int dxdx' \Psi\left(\frac{x+x'}{\sqrt{2}}\right)\delta\left(\frac{-x+x'}{\sqrt{2}}\right)\ket{x}\ket{x'}\nonumber\\
&=&\int dx \Psi(x)\ket{\frac{x}{\sqrt{2}}}\ket{\frac{x}{\sqrt{2}}}. \nonumber
\end{eqnarray} 

The probability $P(x,p)$ for measuring $x,p$ at the two homodyne detections is obtained as the scaled Husimi $\cal{Q}$ function of the coherent state basis $|\mu\rangle\langle \mu|$, where $\mu=x+ip$:
\begin{eqnarray}
P(x,p)&=&\int d^2\alpha~2\mathcal{P}(\alpha) \abs{\Psi(\sqrt{2}x)}^2 \abs{\Psi(\sqrt{2}p)}^2 \nonumber\\
&=&\frac{2}{\pi}\int d^2\alpha~\mathcal{P}(\alpha) e^{-\abs{\alpha - \sqrt{2}\mu}^2} \nonumber\\
&=& 2{\cal{Q}} (\sqrt{2}\mu).
\end{eqnarray}

As a result of the homodyne detections, the state is collapsed to the measurement basis $|x \rangle\langle x | \otimes |p \rangle\langle p |$. However, the obtained probability $P(x,p)$ comes from the coherent state basis measurement acting on the initial state $\hat{\varrho}$. Thus, we expect to collapse the initial state to the coherent state basis after this measurement. To achieve this, we additionally perform a vacuum basis measurement at the end of the scheme. The vacuum basis measurement can be implemented by selecting the event when photodetector 1 (PD1) is not clicked. This conditional state $\hat{\phi}$ becomes the coherent state $|\mu\rangle\langle \mu |$~\cite{Leonhardt93,Lee:2003we}, i.e.,
\begin{equation}\label{eq:xp}
\hat{\phi}=\Tr_2 \left[|0\rangle_2\langle 0| \left(\hat{B}_4|x \rangle_1\langle x |\otimes |p \rangle_2\langle p|\hat{B}_4^{\dagger} \right)\right]=|\mu\rangle_1\langle \mu|, 
\end{equation}
where $\hat{B}_4$ is the operator of beam splitter 4 (BS4). The proof for this equality is shown in Appendix~\ref{eq:EP}.

The sequential measurements of coherent state bases can be realized by performing this $\mathcal{Q}$-function measurement consecutively as depicted in Fig.~\ref{fig:tele}(b). In practical experiments, the zero eigenstates can be replaced by the vacuums highly squeezed in the $x$- or $p$-quadrature direction. Then the practical accuracy of this measurement depends on the squeezing degrees of the ancillary vacuum states.

\section{Conclusions and discussions}
We suggest the operational quasiprobability for the continuous-variable systems (OQCV). It involves the sequential measurements of quadrature variables. The commensurability of our approach enables one to directly compare the OQCV to its classical counterpart, probability distribution, on the same footing. As a classical model, we consider the macrorealistic model, assuming the NSIT and AoT conditions~\cite{Kofler13,Clemente15,Clemente16}. In the classical model, OQCV becomes a joint probability distribution of the sequential measurements. Therefore, the negative values of the OQCV imply the violation of the classical model, i.e., the condition of NSIT or AoT. We show that vacuum, coherent, squeezed vacuum, number, cat and thermal states in low average photon number have negativities in their OQCV. On the other hand, the OQCV function of the bright thermal state converges to the domain of NSIT in the limit of an infinite number of average photons and thus its limit is characterized as classical. We also propose a feasible optical scheme to realize the sequential measurements of quadrature variables.

The commensurate approach can be extended to the scenarios having more than two temporally (or spatially) separated observers sharing quantum systems. We reported such results for discrete quasiprobability in Ref.~\cite{Ryu2013}. It still remains an open problem to clarify differences between the conditions of NSIT and non-negative OQCV as shown in Fig.~\ref{fig:Venn}. 

\acknowledgements
The authors thank J. Sperling for bringing their attention to Ref.~\cite{{Sperling16}}. This research was supported by the National Research Foundation of Korea (NRF) Grant No. 2014R1A2A1A10050117, funded by the MSIP (Ministry of Science, ICT and Future Planning), Korean government. It was also supported by the MSIP (Ministry of Science, ICT and Future Planning), Korea, under the ITRC (Information Technology Research Center) support program (Program No. IITP-2017-2015-0-00385) supervised by the IITP (Institute for Information \& Communications Technology Promotion). J.R. acknowledges that this research is supported by the National Research Foundation, Prime Minister’s Office, Singapore and the Ministry of Education, Singapore under the Research Centres of Excellence programme.

\setcounter{equation}{0}
\renewcommand{\d}[1]{\ensuremath{\operatorname{d}\!{#1}}}
\renewcommand{\thesection}{A\arabic{section}}
\renewcommand{\theequation}{A\arabic{equation}}
\appendix
\section{Characteristic tensor}
\label{eq:character}
We here show that the characteristic tensor $C_{mn}$ is obtained by the expectation value of the outcomes. The expectation of Hermite polynomial moments reads
\begin{eqnarray*}
\langle H_m(x_1)H_n(x_2)\rangle=\int\int dx_1dx_2~P(x_1,x_2)H_m(x_1)H_n(x_2).
\end{eqnarray*}
As we pointed out, the probability distribution can be expanded by Hermite polynomials in the form of Eq.~(\ref{eq:C}). Then by the orthogonality in Eq.~(\ref{eq:ortho}), one has
\begin{widetext}
\begin{eqnarray}
\langle H_m(x_1)H_n(x_2)\rangle &=&
\int\int dx_1dx_2 \left(\sum_{k,l=0}^{\infty}\frac{e^{-\frac{1}{2}x_1^2}e^{-\frac{1}{2}x_2^2}C_{kl}}{(\sqrt{2\pi})^2k!l!}  H_k(x_1)H_l(x_2) \right) H_m(x_1)H_n(x_2)\nonumber\\
&=&\sum_{k,l=0}^{\infty}\delta_{mk}\delta_{nl}C_{kl}=C_{mn}.
\end{eqnarray}
\end{widetext}

\section{Probability representation of the OQCV}
\label{eq:deriving}
\renewcommand{\theequation}{B\arabic{equation}}
To obtain the probability representation in Eq.~(\ref{eq:w}), we expand the distribution $\mathcal{W}(x_1,x_2)$ into the characteristic tensors $\Gamma_{00}$, $\Gamma_{0n}$, $\Gamma_{m0}$, and $\Gamma_{mn}$ in Eq.~(\ref{eq:gamma}): 
\begin{widetext}
\begin{equation}
\begin{aligned}
\mathcal{W}(x_1,x_2)&=e^{-\frac{1}{2}(x_1^2+x_2^2)} \left( \Gamma_{00} + \sum_{m\neq 0}^{\infty} \frac{\Gamma_{m0}}{2\pi m!} H_m(x_1) + \sum_{n\neq 0}^{\infty} \frac{\Gamma_{0n}}{2\pi n!} H_n(x_2) + \sum_{m,n\neq 0}^{\infty} \frac{\Gamma_{mn}}{2\pi m!n!} H_m(x_1)H_n(x_2) \right) \nonumber \\
&= e^{\text{-}\frac{1}{2}(x_1^2+x_2^2)}  + \frac{1}{\sqrt{2\pi}}e^{\text{-}\frac{1}{2}{x^2_2}}\big( P(x_1|M_1) - 1 \big) + \frac{1}{\sqrt{2\pi}}e^{\text{-}\frac{1}{2}{x^2_1}}\big( P(x_2|M_2) - 1 \big)   \nonumber \\
&+ \bigg[ P(x_1,x_2|M_1,M_2)  - \frac{1}{\sqrt{2\pi}}e^{\text{-}\frac{1}{2}{x^2_2}}\big( P(x_1|M_1,M_2)  - 1 \big) - \frac{1}{\sqrt{2\pi}}e^{\text{-}\frac{1}{2}{x^2_1}}\big( P(x_2|M_1,M_2) - 1 \big) - e^{\text{-}\frac{1}{2}(x_1^2+x_2^2)} \bigg] \nonumber \\
&= P(x_1,x_2|M_1,M_2) + \frac{1}{\sqrt{2\pi}}e^{\text{-}\frac{1}{2}{x^2_2}}\bigg( P(x_1|M_1)-P(x_1|M_1,M_2) \bigg) + \frac{1}{\sqrt{2\pi}}e^{\text{-}\frac{1}{2}{x^2_1}}\bigg( P(x_2|M_2)-P(x_2|M_1,M_2) \bigg),
\end{aligned}
\end{equation}
\end{widetext}
where we used $P(x_1|M_1)=e^{-\frac{1}{2}x^2_1}\sum_{m=0}^{\infty}\frac{\Gamma_{m0}}{2\pi m!}H_m(x_1)$ and $P(x_2|M_2)$ similarly.

Equation~(\ref{eq:quadrature}) is derived when the Hermite polynomial of $x_i$ is replaced by the two Hermite polynomials of the continuous-variable pair $(\alpha_r, \alpha_i)$, i.e., $H_m(x_1) \rightarrow H_p(\alpha_r)H_q(\alpha_i)$ and $H_n(x_2) \rightarrow H_r(\beta_r)H_s(\beta_i)$. The observations without measurement $M_{1}$ or $M_{2}$ in Figs.~\ref{fig:setting}(a)--(c) are also distinguished by zero index in the pair $(p,q)=(0,0)$ or $(r,s)=(0,0)$.

\section{Negativity of thermal state}
\label{SEC:APP_C}
\renewcommand{\theequation}{C\arabic{equation}}
We show that the OQCV of the thermal state converges to the joint probability for sequential measurement $M_1, M_2$ in the limit of an infinite number of average photons, $\bar{n}_{\text{th}}$. For the thermal state $\sum_{n=0}^{\infty}{({\bar{n}_{\text{th}}})^{n}}/{(\bar{n}_{\text{th}}+1)^{n+1}}|n\rangle\langle{n}|$, the OQCV is given by
\begin{equation}
\begin{aligned}
\mathcal{W}&(\vec{\alpha},\vec{\beta}) = P(\vec{\alpha},\vec{\beta}|M_1,M_2) \\
&+ \frac{1}{2\pi}e^{-\frac{1}{2}\abs{\alpha}^2}\left[ P(\vec{\beta}|M_2)-P(\vec{\beta}|M_1,M_2) \right],
\end{aligned}
\end{equation}
where the probabilities involved in composing the OQCV are
\begin{eqnarray}
&&P(\vec{\alpha},\vec{\beta}|M_1,M_2)=\frac{e^{-\abs{\alpha}^2/(\bar{n}_{\text{th}}+1)}e^{-\abs{\alpha-\beta}^2}}{\pi^2({\bar{n}_{\text{th}}+1)}}, \nonumber\\
&&P(\vec{\beta}|M_2)=\frac{e^{-\abs{\beta}^2/(\bar{n}_{\text{th}}+1)}}{\pi({\bar{n}_{\text{th}}+1)}},~
P(\vec{\beta}|M_1,M_2)=\frac{e^{-\abs{\beta}^2/(\bar{n}_{\text{th}}+2)}}{\pi(\bar{n}_{\text{th}}+2)}.\nonumber
\end{eqnarray}

We expand the probabilities in power series of $\bar{n}_{\text{th}}$, in the limit of $\bar{n}_{\text{th}} \rightarrow \infty$,
\begin{eqnarray}
&&P(\vec{\alpha},\vec{\beta} | M_1, M_2) \rightarrow \frac{A(\vec{\alpha},\vec{\beta})}{\bar{n}_{\text{th}}}~~\text{and} \nonumber\\
&&\frac{1}{2\pi}{e^{-\frac{1}{2}\abs{\alpha}^2}} \left( P(\vec{\beta}|M_2) - P(\vec{\beta} | M_1, M_2) \right) \rightarrow {\frac{B(\vec{\alpha},\vec{\beta})}{\bar{n}^2_{\text{th}}}},\nonumber
\end{eqnarray}
where $A$ and $B$ are functions depending on the measurement basis choices $\vec{\alpha},\vec{\beta}$. It is clear in the limit of $\bar{n}_\text{th} \rightarrow \infty$ that the condition of no signaling in time holds, as its order $1/\bar{n}^2_\text{th}$ converges to zero more rapidly than $1/\bar{n}_\text{th}$, the order for the joint probability term. In this sense, we say that the OQCV function of the bright thermal state converges to the domain of no signaling in time.


\section{Postmeasurement state in the coherent state}
\renewcommand{\theequation}{D\arabic{equation}}\label{eq:EP}
Here we show that the conditional postmeasurement state in Eq.~(\ref{eq:xp}) turns out to be the coherent state. The state in Eq.~(\ref{eq:entangle}) collapses to the measurement basis $|x \rangle_1\langle x |\otimes |p \rangle_2\langle p|$ when quadrature variable $x,p$ measurements are performed. It was shown in Ref.~\cite{Lee:2003we} that these quadrature eigenstates and beam splitter operation result in the state
\begin{equation}
\hat{B}|x \rangle_1\langle x |\otimes |p \rangle_2\langle p|\hat{B}^{\dagger}=\hat{D}_1(x+ip)|\Phi\rangle_{12}\langle\Phi|\hat{D}_1^{\dagger}(x+ip),
\end{equation}
where the $\hat{D}$ is a displacement operator and $\ket{\Phi}$ is the (unnormalized) maximally entangled state. By the acting vacuum basis measurement at mode 2, then the state becomes the coherent state $|x+ip\rangle\langle x+ip|$,
\begin{eqnarray}
&&\Tr_2\left[|0\rangle_2\langle0|~\hat{D}_1(x+ip)|\Phi\rangle_{12}\langle\Phi|\hat{D}_1^{\dagger}(x+ip)\right] \nonumber\\ 
&=& \int d^2 \eta  ~e^{-(x+ip)\eta^*+(x+ip)^*\eta} \int d^2\eta~\hat{D}^{\dagger}(\eta) \otimes \langle0|\hat{D}^{\dagger}(\eta^*)|0\rangle \nonumber\\
&=& \int d^2 \eta ~e^{-\frac{1}{2}\abs{\eta}^2-(x+ip)\eta^*+(x+ip)^*\eta} ~\hat{D}^{\dagger}_1(\eta) \nonumber\\
&=& \int d^2 \eta ~\Tr \big[|x+ip\rangle\langle x+ip| \hat{D}(\eta)\big] \hat{D}^{\dagger}(\eta) \nonumber\\
&=& |x+ip\rangle\langle x+ip|.
\end{eqnarray}
We use the relation for a maximally entangled state $|\Phi\rangle_{12}\langle\Phi|=\int d^2\eta~\hat{D}_1^{\dagger}(\eta)\otimes\hat{D}_2^{\dagger}(\eta^*)$ in Ref.~\cite{Lee:2003we}. This result is equivalent to the derivation in Ref.~\cite{Leonhardt93}.


\end{document}